\documentclass[aps,prl,superscriptaddress,10pt]{revtex4-2}
\usepackage{amsmath}	

\usepackage{graphicx}	
\usepackage{color}
\usepackage{gensymb}
\usepackage{braket}
\usepackage{hyperref}
\usepackage{upgreek}

\begin{document}
\Large

\title{\Large Exciton-polaritons in a monolayer semiconductor coupled to van der Waals dielectric nanoantennas on a metallic mirror}

\author{Yadong Wang}
\email{yadong.wang@sheffield.ac.uk}
\affiliation{School of Mathematical and Physical Sciences, University of Sheffield, Sheffield S3 7RH, UK}

\author{Charalambos Louca}
\affiliation{School of Mathematical and Physical Sciences, University of Sheffield, Sheffield S3 7RH, UK}

\author{Sam Randerson}
\affiliation{School of Mathematical and Physical Sciences, University of Sheffield, Sheffield S3 7RH, UK}
\author{Xuerong Hu}
\affiliation{School of Mathematical and Physical Sciences, University of Sheffield, Sheffield S3 7RH, UK}

\author{Panaiot G. Zotev}
\affiliation{School of Mathematical and Physical Sciences, University of Sheffield, Sheffield S3 7RH, UK}
\author{Oscar Palma Chaundler}
\affiliation{School of Mathematical and Physical Sciences, University of Sheffield, Sheffield S3 7RH, UK}
\author{Paul Bouteyre}
\affiliation{School of Mathematical and Physical Sciences, University of Sheffield, Sheffield S3 7RH, UK}
\author{Casey K. Cheung}
\affiliation{Department of Physics and Astronomy, The University of Manchester, Oxford Road, Manchester, M13 9PL, UK}
\affiliation{National Graphene Institute, The University of Manchester, Oxford Road, Manchester, M13 9PL, UK}
\author{Roman Gorbachev}
\affiliation{Department of Physics and Astronomy, The University of Manchester, Oxford Road, Manchester, M13 9PL, UK}
\affiliation{National Graphene Institute, The University of Manchester, Oxford Road, Manchester, M13 9PL, UK}
\author{Yue Wang}
\affiliation{School of Physics, Engineering and Technology, University of York, York, YO10 5DD, UK}
\author{Alexander I. Tartakovskii}
\email{a.tartakovskii@sheffield.ac.uk}
\affiliation{School of Mathematical and Physical Sciences, University of Sheffield, Sheffield S3 7RH, UK}


\begin{abstract}

\end{abstract}

\maketitle

\textbf{Polaritons in nanophotonic structures have attracted long-standing interest owing to their fundamental importance and potential for applications in nonlinear and quantum optics. Nanoantennas (NAs) made from high refractive index dielectrics offer a suitable platform for polariton physics thanks to the strongly confined optical Mie resonances and low optical losses in contrast to metallic NAs. 
However, Mie modes are mainly confined within the NA, making inefficient their coupling with excitons in materials deposited externally. Here, we overcome this limitation by using a high-refractive index van der Waals material WS$_2$, which allows straightforward fabrication of NAs on gold. The combination of a 27 nm tall WS$_2$ NA and a gold substrate enables strong modification of the Mie mode distribution and field enhancement inside and in the vicinity of the NA. This allows observation of room-temperature Mie-polaritons (with a Rabi splitting above 80 meV) arising from the strong coupling between Mie modes and the exciton in a monolayer WSe$_2$ placed on WS$_2$/gold NAs. We demonstrate strong nonlinearity of Mie-polaritons, one order of magnitude higher than for excitons in monolayer WSe$_2$ on gold. Our results highlight applicability of van der Waals materials for the realisation of hybrid dielectric-metallic nanophotonics for the study of the strong light-matter interaction.}

\section{Introduction}\label{sec1}

The strong light-matter interaction underlies a wide range of phenomena and applications including Bose-Einstein condensation in the solid state \cite{Byrnes2014}, polariton superfluidity \cite{Lerario2017}, polariton lasing \cite{Christopoulos2007}, low-energy switching \cite{Ballarini2013,Dreismann2016} and optical parametric amplification \cite{Savvidis2000}. Compact optical cavities can be used to enhance the light-matter interaction on the nanoscale and thus plasmonic cavities have been widely studied \cite{Chikkaraddy2016,Benz2016,Kleemann2017}.  Whilst metallic nanocavities have attracted strong interest, high losses in metals still limit their applicability \cite{Baffou2013}. Recently, high-refractive-index dielectric nanoantennas made mostly from traditional photonic materials such as group IV (Si, Ge) and III-V (GaAs, GaP etc) have emerged as an attractive platform for nanophotonics exploiting highly confined optical modes referred to as Mie resonances, which are confined in structures with dimensions comparable to the light wavelength \cite{mie1908beitrage,Caldarola2015,Kuznetsov2016,Cambiasso2017}. Linear and nonlinear optical regimes as well as both individual structures and arrays (or metasurfaces) have been explored \cite{Kuznetsov2016}.

In the context of all-dielectric nanophotonics, layered transition metal dichalcogenides (TMDs) present an attractive extension of the pool of available materials thanks to their high refractive index and low absorption in the visible and near-infrared \cite{Verre2019,Green2020,Busschaert2020,Ermolaev2021,Nauman2021,Munkhbat2022a,Munkhbat2022b,Zotev2022,Popkova2022,Zotev2023}. As an example, in the visible wavelength range, bulk WS$_2$ has a higher refractive index ($n>$4) than gallium phosphide ($n \approx$3.5), silicon ($n \approx$ 3.8), and silicon nitride ($n \approx$ 2.1). TMDs and a very wide class of layered materials to which they belong are often referred to as van der Waals crystals after the forces holding the atomic planes of the crystal together \cite{Geim2013}. Thanks to the van der Waals forces, flakes of such materials can be exfoliated from bulk crystals by an adhesive tape and can be consequently easily attached to a wide range of surfaces allowing a new degree of flexibility in fabrication, e.g. through high-precision flake positioning, and a wide choice of suitable substrates \cite{Frisenda2018}. These advantages promote van der Waals materials as promising candidates for fabrication of nanophotonic structures, which has led to recent demonstrations of TMD nanoantennas and metasurfaces \cite{Verre2019,Green2020,Busschaert2020,Ermolaev2021,Nauman2021,Munkhbat2022a,Munkhbat2022b,Zotev2022,Popkova2022,Zotev2023}. 

Recent experiments reporting strong coupling in dielectric systems relied on collective resonances from periodic arrays of nanostructures, as the principal way for achieving sharp resonances \cite{Castellanos2020,Wang2020,Qin2021,Zong2021,Kumar2022,Weber2023,Cho2023,Lee2023,Bouteyre2025}. In TMD-based structures, the focus has been on the strong coupling with bulk excitons both in periodic structures \cite{Kumar2022,Weber2023,Cho2023,Lee2023,Bouteyre2025} and in single dielectric nanoantennas \cite{Verre2019,Zotev2022,Zotev2023}. The realization of Mie-polaritons in individual dielectric nanoparticles coupled with 
excitons in monolayer TMDs have been considered theoretically \cite{Lepeshov2018,Tserkezis2018}. However, these models use hard-to-realize structures made of a spherical dielectric "core" covered with a TMD "shell" \cite{Lepeshov2018,Tserkezis2018}. 

Here, we realize Mie-polaritons in a semiconducting TMD monolayer WSe$_2$. For this observation, we rely on Mie-resonances in WS$_2$ NAs fabricated on a gold substrate. The strong reflection from the gold substrate considerably increases the confinement of the Mie modes and thus enhances their quality factors. The presence of gold thus modifies the mode distribution, and in particular leads to a large electric field intensity surrounding the top surface of the NAs, where we place a WSe$_2$ monolayer. We observe a large Rabi splitting between the upper (UB) and lower (LP) polariton branches of $\sim$78 meV in reflectance spectra and $\sim$86 meV using dark-field spectroscopy. Furthermore, by increasing the excitation density, we found that Mie polaritons exhibit an order of magnitude higher nonlinearity than excitons in monolayer WSe$_2$ on gold.

\begin{figure}[t]%
\includegraphics[width=1\textwidth]{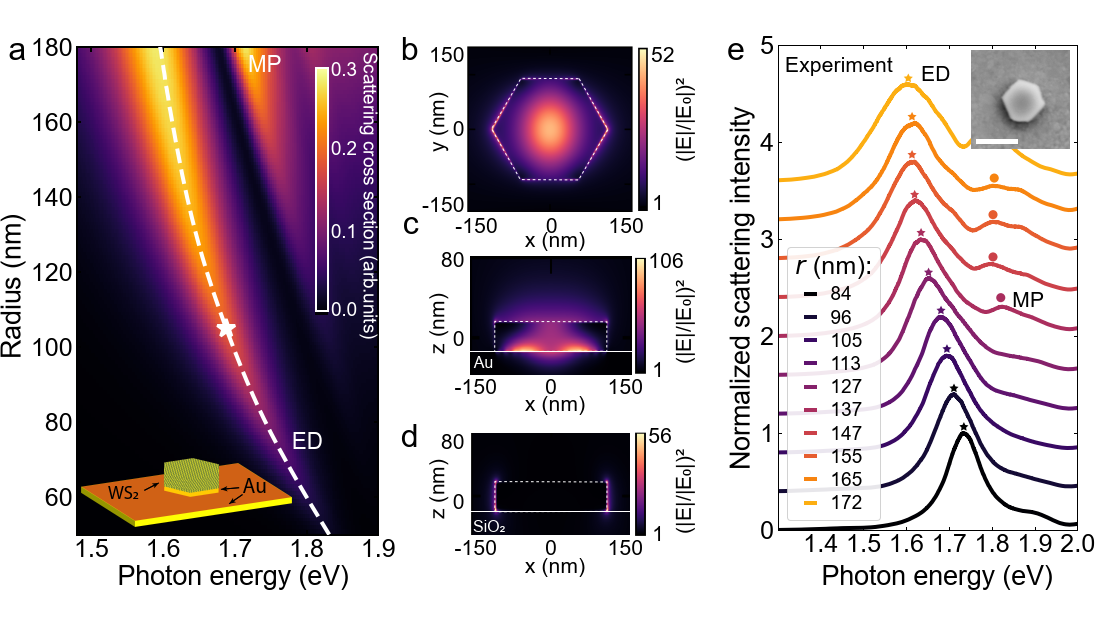}
\caption{\large \textbf{Mie resonances in multilayer WS$_2$ nanoantennas (NAs).} \textbf{a,} Simulated light scattering cross-section of a WS$_2$/Au NA on gold with radii ranging from 50 nm to 180 nm. The NAs are 27 nm thick WS$_2$ on top of a 3 nm gold pedestal arising due to a slight over-etching in the fabricated structures. The white star indicates the mode for which further details are shown in Fig. 1b and 1c. ED and MP labels the electric dipole and Mie-plasmonic modes (see details in text and Ref.\cite{Randerson2024}). Inset: Schematic of a WS$_2$ NA/Au hybrid structure.  The lateral (\textbf{b}) and vertical (\textbf{c}) cross-sections through the centre of the NA showing the electric-field distributions of NAs with a radius of 105 nm at a photon energy of 1.69 eV. \textbf{d,} Electric field distribution in a WS$_2$ NA placed on SiO$_2$, where the electric field is simulated for a WS$_2$ NA with the thickness of 30 nm and radius 105 nm at a photon energy of 1.94 eV corresponding to an ED mode (see further details in Fig. S3 in the SI). \textbf{e,} Experimentally measured dark-field (DF) spectra of the WS$_2$ NAs with radii ranging from 84 nm to 172 nm. Inset: SEM image of the NA. Scale bar: 200 nm.}
\label{fig1}
\end{figure}

\section{Results}\label{sec2}

We first perform finite-difference time-domain (FDTD) simulations of WS$_2$ NAs on a gold substrate, illuminated by linearly-polarized light at normal incidence (see the schematic of the structure in the inset of Fig. \ref{fig1}a). We use a hexagonal cross-section in accordance with the fabricated NAs \cite{Zotev2023}. NAs with a total height of 30 nm (consisting of 27 nm WS$_2$ and 3 nm Au) and radii in the range of 50 nm to 180 nm, reproduce well the experimental results (see Fig. S1 in the Supplementary Information, SI). Note that during the ion etching process, the Au substrate has also been etched by around 3 nm. The refractive index of WS$_2$ used in these simulations is taken from previous measurements using spectroscopic micro-ellipsometry \cite{Zotev2023}. Figure \ref{fig1}a shows the simulated scattering cross-section of WS$_2$ NAs with radii ($r$) ranging from 50 to 180 nm. Two distinct peaks are observed. The low-energy peak, tunable from 1.83 to 1.59 eV with increasing radius (see Figs. S1 and S2 in the SI), is assigned to the electric-dipole resonance, labelled as ED \cite{Randerson2024}. Its linewidth notably increases as it red-shifts with the increasing NA radius as found from Lorentzian fits. The linewidth changes from $\sim$52 meV (for $r=50$ nm) to $\sim$168 meV ($r=180$ nm). These correspond to $Q$ factors of $\sim$35.3 and $\sim$9.5 respectively (see Fig.S2 in the SI). The second peak at higher energy in Fig. \ref{fig1}a appears gradually for a larger radius, and is attributed to the Mie-plasmonic mode (MP), arising from the hybridization of Mie modes and surface plasmon polaritons at the TMD-gold interface. More detailed analysis of such Mie-plasmonic modes can be found in our recent work \cite{Randerson2024}. 

Here, we focus on the lower-energy electric-dipole mode. We choose a specific NA with a radius of 105 nm (star in Fig. \ref{fig1}a) to analyze its near-field behaviour. The electric field distributions for the photon energy  1.69 eV plotted in the form $|E|^2/|E_0|^2$ ($E$ and $E_0$ is the electric field inside the NA and in the incident wave, respectively) in the $x-y$ and $x-z$ planes are displayed in Fig. \ref{fig1}b and Fig. \ref{fig1}c, respectively.  The distribution in the $x-y$ plane exhibits a shape corresponding to an electric-dipole mode.  As shown in the electric-field profile in the $x-z$ plane, the ED mode is strongly influenced by the presence of the gold substrate \cite{Randerson2024}. It is notable that due to the presence of the gold substrate, high electric field intensities can be found above the top surface of the NA. We believe that it is this feature occurring thanks to the use of the thin WS$_2$ and gold substrate which allows the observation of strong coupling with a monolayer attached on top of the NA as reported below. We also simulated WS$_2$ NAs on a SiO$_2$ substrate (see Fig. S3 in the SI) for comparison. Figure \ref{fig1}d shows the $x-z$ plane of the electric field distribution ($|E|^2/|E_0|^2$) in the middle of the NA on SiO$_2$ substrates at 1.94 eV. The electric field inside of the NA is weak with a maximum $|E|^2/|E_0|^2$=0.34. Compared to that on Au substrate where $|E|^2/|E_0|^2$ reaches 44.2, this yields an enhancement due to the presence of the gold substrate of $\sim$130.

\begin{figure}[t]
\centering
\includegraphics[width=0.8\textwidth]{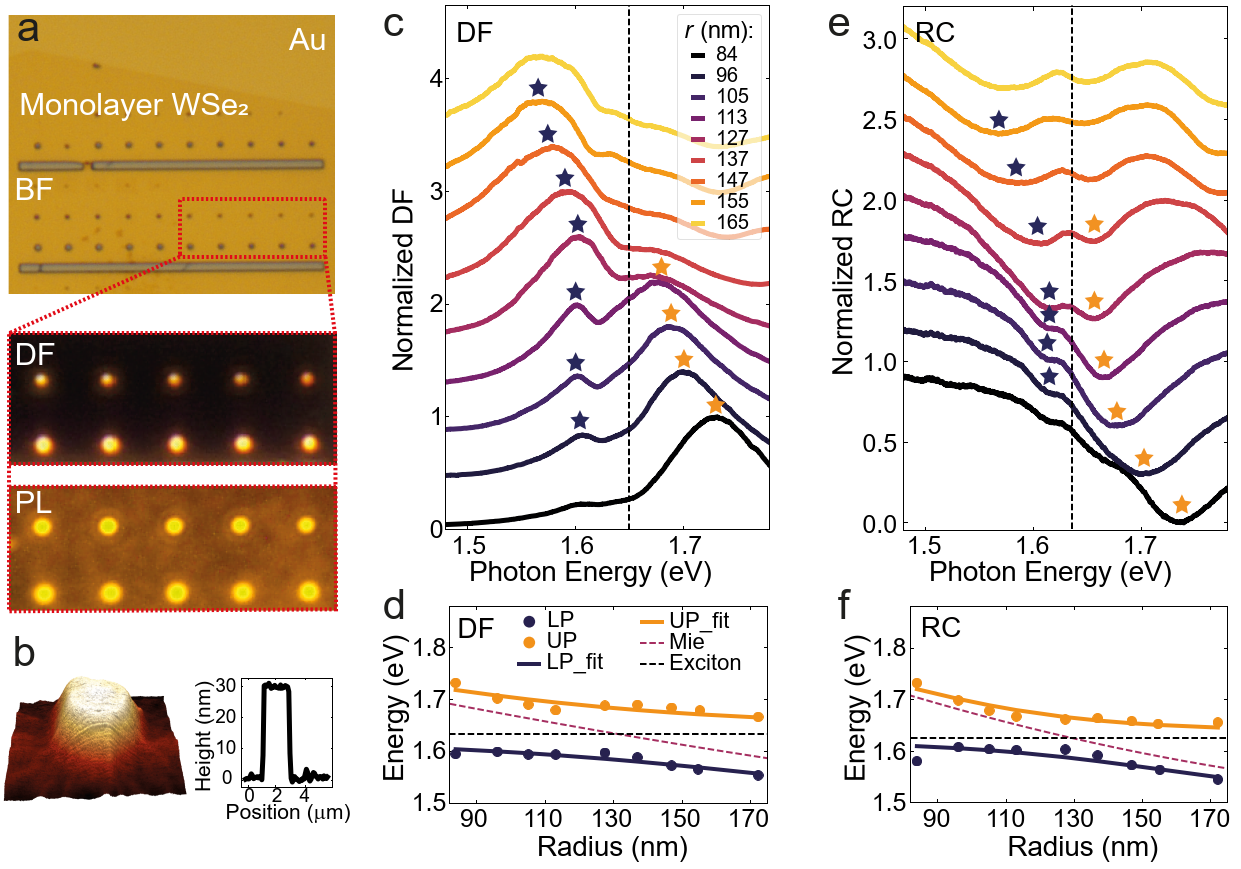}
\caption{\large{\bf{Strong coupling between the Mie resonance in hybrid WS$_2$/gold NAs and exciton in a monolayer WSe$_2$ placed on top of the NAs.}} \textbf{a,} Optical image of an array of WS$_2$ NA/Au hybrid structures. The enlarged area shows dark field (DF, top) and photoluminescence (PL, bottom) images of the NAs within the red dashed box. \textbf{b,} AFM image (left) and cross-section displaying the height profile (right) of the monolayer WSe$_2$ covering a WS$_2$/gold NA. \textbf{c,} Experimentally measured dark-field (DF) and \textbf{e,} reflectance contrast (RC) spectra of 1L WSe$_2$/NAs with the radii in the range of 84-165 nm. The blue and orange stars denote the lower and upper polariton branches, and the dashed black line marks the 1L WSe$_2$ exciton energy. \textbf{d} and \textbf{f,}  Peak positions in the DF and RC spectra from Fig. \ref{polariton_exp}c and e (circles). Dashed lines show the uncoupled Mie mode and WSe$_2$ exciton energies. Solid lines denote the fitting of the experimental dependences with the coupled-oscillator model, yielding a Rabi splitting of $\Omega$ = 86$\pm$5 meV  for the DF and $\Omega$ = 78$\pm$6 meV for the RC spectra, respectively.}\label{polariton_exp}
\end{figure}

For our experiment, we exfoliate a 27-nm thick WS$_2$ flake on a 150-nm thick gold film that was deposited on a SiO$_2$/Si silicon substrate by electron beam evaporation. The WS$_2$ NAs were fabricated by electron beam lithography followed by reactive ion etching. The scanning electron microscopy (SEM) image of WS$_2$ NAs as shown in the inset of Fig. \ref{fig1}e exhibits a typical hexagonal-like shape in accordance with the crystal structure of WS$_2$, highlighting the effect of the selective chemical etching \cite{Zotev2022,Zotev2023,Randerson2024,Munkhbat2020}. The radii of these NAs are identified from the SEM images. The NAs of different sizes were fabricated in arrays (see images in Fig. \ref{polariton_exp}a) so that consequently all of them can be covered by a monolayer WSe$_2$ flake. Individual structures were separated by a relatively large distance of 3 $\mu$m so that individual NAs can be probed. We do not observe any collective excitation modes corresponding to the whole array. 

The light scattering measurements were performed on individual NAs using a commercial optical microscope in the dark-field mode (see details in the Methods section). Figure \ref{fig1}e shows the dark-field spectra of NAs with different radii. The strongest peak at around 1.7 eV, corresponding to the ED mode, red-shifts with increasing radius, which agrees well with the simulation results. The linewidths range from 80 meV to 130 meV fitted by a Lorentz function, corresponding to a $Q$ factor of 10 and 17 respectively, slightly lower than the simulation results. The observed higher energy peaks correspond to the Mie-plasmonic (MP) mode arising in the NAs with radii larger than 127 nm. Note, that here we present spectra for the NAs themselves, which consequently were covered with a monolayer WSe$_2$ and measured again, thus allowing us to measure the changes that the monolayer WSe$_2$ induces in the spectra of each NA. 

After the characterization of the WS$_2$ NAs, an exfoliated monolayer WSe$_2$ was placed onto the WS$_2$ antennas using a dry transfer method. As shown in Fig. \ref{polariton_exp}a, the monolayer WSe$_2$ is visible in the optical microscope image. The red dashed rectangular region is highlighted to show the coupling between the monolayer WSe$_2$ and NAs in the dark-field (DF) and photoluminescence (PL) image (see details in methods) \cite{alexeev2017imaging}. The DF image shows that the monolayer WSe$_2$ was transferred onto the NAs without wrinkles. In the PL image, the monolayer WSe$_2$ emits bright PL on the NAs sites, whereas the monolayer WSe$_2$ on the Au substrate shows a much lower PL intensity due to quenching by the gold \cite{hasz2022tip}. More details on the PL emission and photonic enhancement are reported in Fig. S4 in the SI. The atomic force microscopy (AFM) scan of one of the NAs further shows that the transferred, atomically thin layer of WSe$_2$ closely follows the shape of the NA (Fig. \ref{polariton_exp}b).

We carry out DF measurements for a range of NAs with different radii to explore how the spectrum changes as the Mie mode is tuned through the resonance with the WSe$_2$ exciton. The results are shown in Fig. \ref{polariton_exp}c, where a clear anti-crossing behaviour is observed (see detailed comparison of the DF spectra with and without the WSe$_2$ monolayer in Fig. S5 in the SI). The blue and orange stars in Fig. \ref{polariton_exp}c show the tuning of the Mie-polariton branches and the vertical dashed black line denotes the position of the exciton in the WSe$_2$ monolayer. We extract the peak positions and subsequently fit them using a coupled oscillator model as shown in Fig. \ref{polariton_exp}d (see the discussion in the SI). From the fitting, we obtain a Rabi splitting of $\hbar\Omega= 86\pm5$ meV. By considering the linewidths of the exciton ($\gamma_0$ = 56$\pm$3 meV) and Mie resonance ($\gamma_{Mie}$ = 92$\pm$2 meV) extracted from the RC measurements in monolayer WSe$_2$ and the DF measurements on the uncoupled NAs (close to the anti-crossing), the coupling strength is larger than half of the sum of the individual damping rates, i.e. $\hbar\Omega_R > (\gamma_0+\gamma_{Mie})/2$, indicating that the system is in the strong coupling regime. The results of the fitting are further presented in Fig. \ref{polariton_exp}d, where the symbols show the peak positions measured experimentally, the solid lines show the modelled behaviour of Mie-polaritons, and the dashed lines show the spectral positions of the uncoupled exciton and Mie-resonance. 

Further to our observation of the strong coupling using DF spectroscopy, we repeated the measurements for the monolayer WSe$_2$  on different NAs using reflectance contrast (RC). The formula for RC intensity is $R_{NA}/R_{Au}$, where $R_{NA}$ and $R_{Au}$ are the reflectance intensities from the NA and the bare gold substrate, respectively. As shown in Fig. \ref{polariton_exp}e, two dips appear in the RC spectra showing clear evidence of an anti-crossing and thus strong coupling. Similarly to the procedure we followed with the DF spectra, we fitted each curve using two Lorentzians and then applied the coupled oscillator model as shown in Fig. \ref{polariton_exp}f to extract the Rabi splitting of $\hbar\mathrm{\Omega}_{\mathrm{R}}$ = 78$\pm$6 meV, indicating the system is still in the strong coupling regime.

\begin{figure}[t]
\centering
\includegraphics[width=1\textwidth]{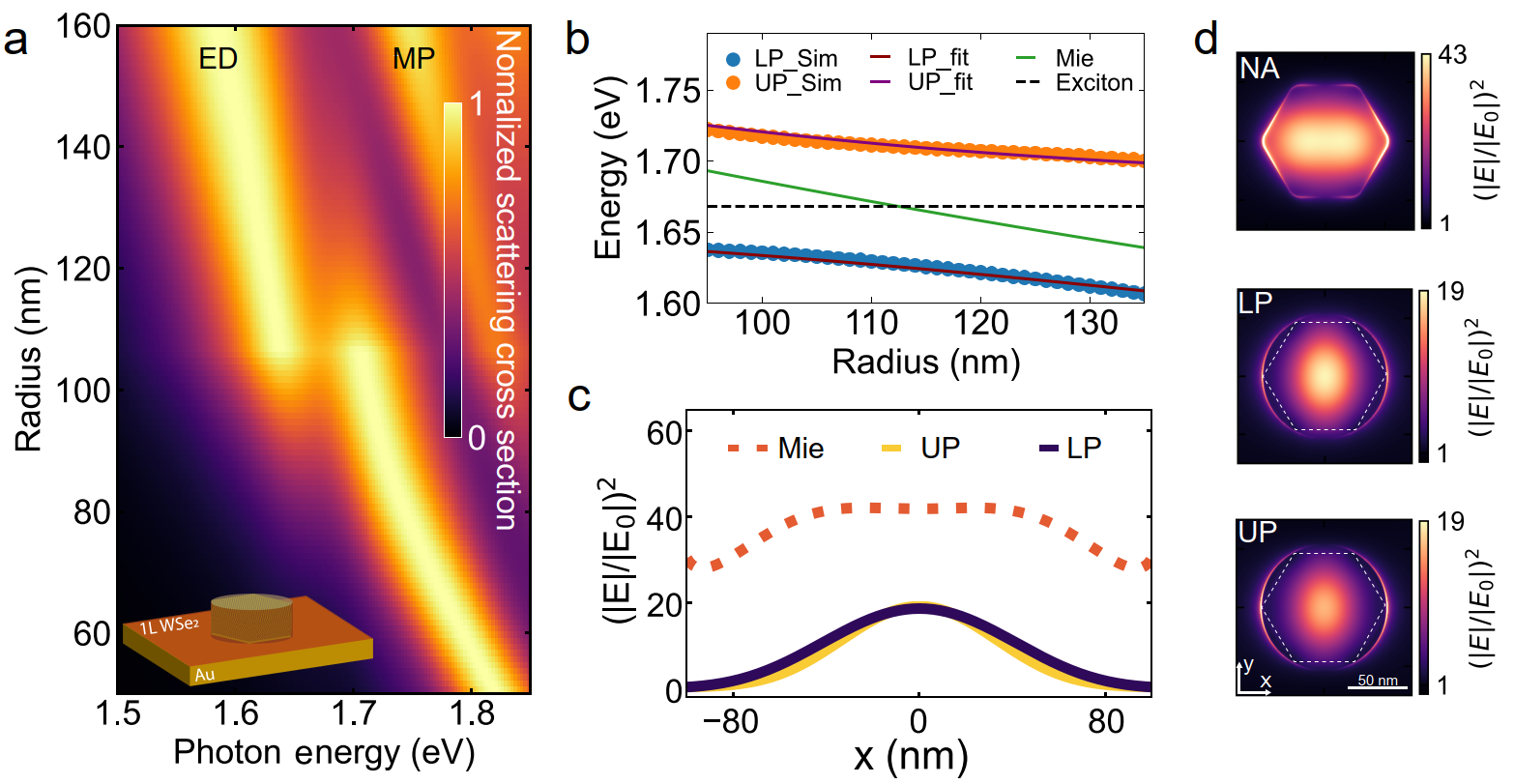}
\caption{\large {\bf Simulation of Mie-polaritons in a monolayer WSe$_2$ enveloped WS$_2$ NA/Au hybrid structure.} \textbf{a,} Normalized scattering intensity of the
monolayer WSe$_2$/WS$_2$ NAs on an Au substrate. The NA height is composed of 27 nm of WS$_2$ and 3 nm gold. Radii range from 50 nm to 160 nm. Inset: schematic of the simulated structure containing a monolayer WSe$_2$ in a form of a hollow cylinder wrapped around a WS$_2$/gold NA on a gold substrate. \textbf{b,} Peak positions of the resonances in Fig. \ref{polariton_sim}a obtained with Lorentz fitting. Solid black lines show the fitting obtained using the coupled-oscillator model yielding a Rabi splitting of $\Omega$ = 85.4 $\pm$ 0.47 meV. Green line represents the energies of the uncoupled Mie resonances. Dashed line shows the exciton energy in a WSe$_2$ monolayer. \textbf{c,}  The relative electric field intensity in the $x-y$ plane at the top of a WS$_2$ NA (dashed orange curve) and in a monolayer-wrapped WS$_2$ NA for the lower (LP, dark blue) and upper (UP, yellow) polariton branches. \textbf{d,} The relative electric field intensity in the $x-y$ plane at the top of an NA for a WS$_2$ NA without a monolayer WSe$_2$ (top panel) and  a WS$_2$ NA with a monolayer WSe$_2$ with the intensity distribution for the lower (middle) and upper (bottom) polariton branches. 
}\label{polariton_sim}
\end{figure}


We further modelled the observed strong coupling in FDTD. Similar to earlier models \cite{lepeshov2018tunable,tserkezis2018mie,wang2019resonance} and in agreement with the experimental realisation of the WSe$_2$-NA system as observed in AFM, we consider a monolayer WSe$_2$ wrapping a WS$_2$ NA on all sides except the bottom facet adjacent to the gold substrate. We model the 1L WSe$_2$ assuming a cylindrical shape as shown in the insert of Fig. \ref{polariton_sim}a, and use the refractive index for 1L WSe$_2$ from Ref. \cite{2019RN406}. Figure \ref{polariton_sim}a shows the normalized scattering intensity with the NA radius ranging from 50 nm to 160 nm, where a clear Rabi splitting has been observed. We fit the spectra simulated for antennas of different radii in Fig. \ref{polariton_sim}a with Lorentz peaks and obtain peak positions as shown with symbols in Fig. \ref{polariton_sim}b. The semi-classical coupled oscillator model is then used to fit the data and extract a Rabi splitting of $\hbar\Omega_R$ = 85.4$\pm$0.47 meV, with the exciton at 1.668 eV. This Rabi splitting is in a good agreement with the value obtained from our dark-field measurements. For comparison, we also simulate a WSe$_2$ monolayer enveloping WS$_2$ NAs on a SiO$_2$ substrate (instead of gold), which shows no evidence of the strong coupling between the exciton in WSe$_2$ and Mie resonances in WS$_2$ NAs (see details in Fig. S6).

Our simulations show that the electric-field distributions of the Mie-resonance in the WS$_2$/gold NA and the Mie-polariton states in the coupled system differ significantly as shown in Fig. \ref{polariton_sim}c. The mode in the WS$_2$/gold NA (dotted orange curve) is much wider than those of the lower polariton (dark blue curve, energy at 1.64 eV) and upper polariton (yellow curve, energy at 1.72 eV) profiles for the coupled system. Furthermore, the maximum of the relative electric-field intensity of the ED mode of the WS$_2$/gold NA ($|E|^2/|E_0|^2$ = 41.8 at $x$ = 0) is almost twice that of the individual Mie-polariton LP ($|E|^2/|E_0|^2$ = 18.6) and UP ($|E|^2/|E_0|^2$ = 19.1) modes, showing that the energy contained in the ED mode is now split between the two polariton states. Figure 3d shows the corresponding electric field distributions in the $x-y$ plane. The electric fields of the polariton modes are more confined in the middle of the NA with a high intensity also observed in the monolayer.

\section{Nonlinearity of Mie-polaritons}\label{sec3}

\begin{figure}[t]
\centering
\includegraphics[width=1\textwidth]{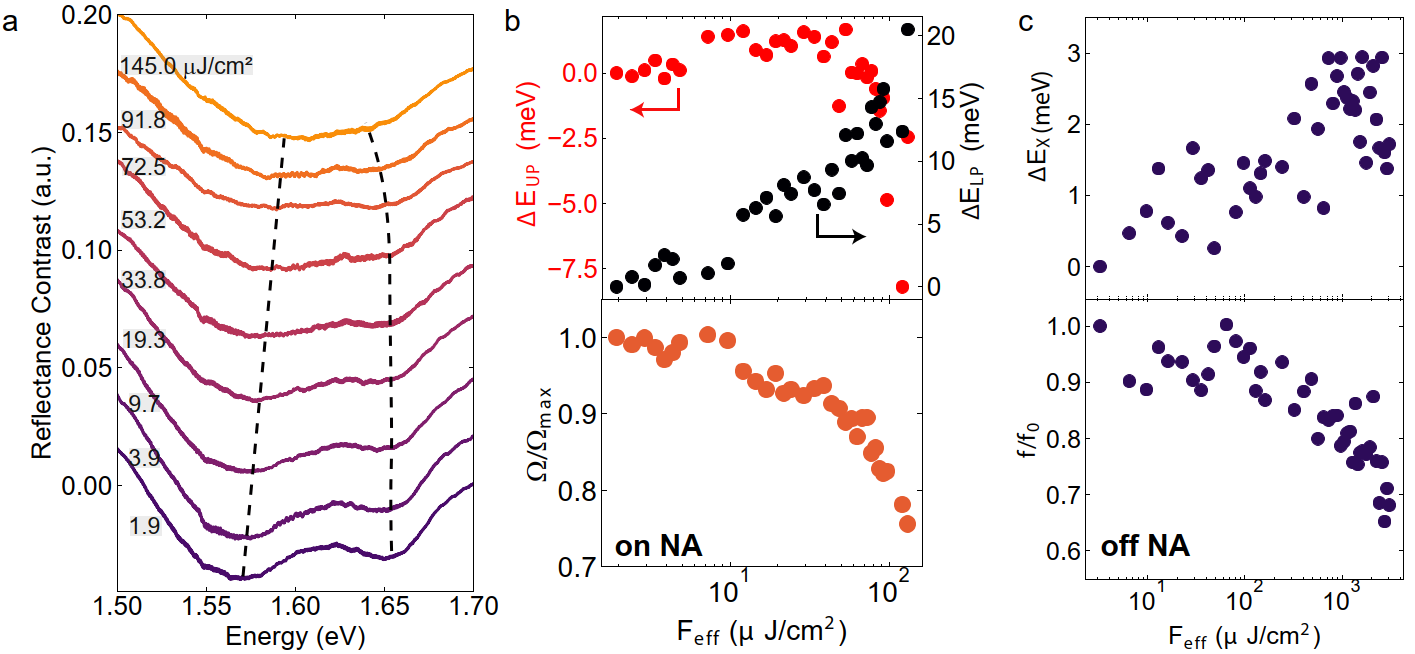}
\caption{\large {\bf Nonlinearity of Mie-polaritons.} \textbf{a,} RC spectra measured using a pulsed laser with fluences ranging from 2 to 145 $\mu$J/cm$^{2}$ on the area of the NA. Power dependence was performed at the NA hosting a resonance with the smallest detuning from the neutral exciton. Dashed lines are a guide to the eye. \textbf{b,} Energy shifts of each polariton branch (upper panel) and normalized Rabi splitting $\Omega$/$\Omega_{max}$  for Mie-polaritons as a function of the effective fluence on the NA area. $\Omega$ is normalised by the maximum Rabi splitting $\Omega_{max}$ observed at the lowest power.  \textbf{c,} The energy shift (upper panel) and normalized integrated intensity (lower panel) of the exciton peak in WSe$_{2}$ monolayer on gold (i.e. outside the NA) as a function of the laser fluence. The normalized intensity is presented as a ratio of the oscillator strengths $f/f_0$ with the maximum $f_0$ observed for the lowest fluence.  
}\label{pdep_polariton} 
\end{figure}

In this part we report a fluence dependence of Mie-polariton RC spectra measured on an NA with the smallest detuning between the Mie-resonance and WSe$_2$ exciton (close to the anti-crossing energy). The Mie-polariton modes were excited with a pulsed ($\sim$20 ps) broadband super-continuum laser with the wavelengths ranging from 650 nm to 850 nm and a repetition rate of 152 kHz. The laser fluence, $F_{eff}$, was varied from 2 to 145 $\mu$J/cm$^2$, corresponding to the incident average power range of 12 nW to 900 nW. Fig. \ref{pdep_polariton}a shows the measured fluence-dependent RC spectra, where we observe a reduction in the energy splitting ($\Omega$) between the upper and lower polariton branches. At high fluences the LP and UP peaks cannot be resolved, indicating the collapse of the strong-coupling regime. This nonlinear behaviour is fully reversible, without any sign of damage of the NA or the WSe$_2$ monolayer. Similarly to the procedure used for Fig. \ref{polariton_exp}, we fit the spectra with two Lorentzians, and plot their shifts as a function of the fluence on the NA area Fig. \ref{pdep_polariton}b (see Fig. S7 in the SI for the details of the fitting).  The maximum energy redshift of the upper polariton ($\Delta E_{UP}$) is found to be 7.5 meV, whilst the lower polariton ($\Delta E_{LP}$) exhibits a larger blueshift of $\sim$20 meV, as displayed in Fig. \ref{pdep_polariton}b. 

The corresponding normalized Rabi splittings $\Omega$/$\Omega_{max}$, where $\Omega_{max}$ is measured at low ﬂuence, are shown in the lower panel of Fig. \ref{pdep_polariton}b as a function of the fluence. We observe $\Omega$/$\Omega_{max}$ reduction down to $\approx 0.7$ for a fluence of 145 $\mu$J/cm$^2$. On the areas of the sample without NAs we observe that the uncoupled excitons exhibit significantly less nonlinearity, as shown in Fig. \ref{pdep_polariton}c, where in the range of the laser fluences reported in Fig. \ref{pdep_polariton}a the exciton shift below 2 meV and negligible variation of the integrated intensity of the exciton peak ($f/f_0$) in the RC spectra are observed. Here, we use the integrated intensity of the exciton in RC as a measure of the oscillator strength $f$, with $f_0$ being the oscillator strength at low excitation powers. Considering $\Omega/\Omega_{max} \propto \sqrt{f/f_0}$, the observed Rabi splitting ratio reduction to $\approx$ 0.7 would correspond to the effective exciton oscillator strength $f$ reduction to $\approx$ 0.49$f_0$, corresponding to at least an order of magnitude stronger variation than in monolayer excitons in the same sample. 

\section{Conclusion}\label{sec4}
 
 In summary, we present the experimental demonstration of room temperature Mie-polaritons in TMD monolayer coupled to hybrid dielectric-metal nanophotonic structures based on quasi-bulk van der Waals materials. We utilise the high refractive index of van der Waals bulk TMDs to reduce the volume of the confined optical modes in nanoantennas made from WS$_2$. We then use the gold mirror below the nanoantennas to increase the Q factor of the modes (compared with WS$_2$ on SiO$_2$) and to suitably modify their electric field distribution thus allowing the coupling of the exciton in the externally placed monolayer semiconductor WSe$_2$ to the confined optical modes. This allows us to achieve strong light-matter coupling in this hybrid system leading to observation of highly confined monolayer-TMD Mie-polaritons. We show that the monolayer-TMD Mie-polaritons are at least an order of magnitude more nonlinear than excitons in the monolayer WSe$_2$, with the Mie-polariton energy state shifts of up to 20 meV observed at the resonant laser fluences of 145 $\mu$J/cm $^{2}$ as opposed to the shifts below 2 meV in the WSe$_2$ monolayer attached on gold in the same device.


\clearpage

\section{Methods}\label{sec5}

\textbf{Sample fabrication}: Thin-layer WS$_2$ and monolayer WSe$_2$ were mechanically exfoliated from bulk crystals (HQ Graphene) by using a low-adhesion tape. The WS$_2$ flakes were exfoliated directly onto a gold film, which was deposited on a silicon chip covered with 10-nm titanium as an adhesion layer, by using electron beam evaporation. WS$_2$ flakes with uniform thickness and lateral dimensions exceeding 20 microns were selected for further nano-fabrication steps using a electron lithography process and reactive ion etching following the procedure in Ref. \cite{Zotev2023}. The monolayer WSe$_2$ was obtained by exfoliating form the bulk crystal onto a PDMS stamp followed by transferring onto the prepared WS$_2$ NAs using a micro-manipulation stage.

\textbf{Bright-field spectroscopy}: Optical spectroscopy in a bright field configuration was carried out using a commercial Nikon LV150N microscope. The white light from a tungsten halogen lamp was guided into a 50$\times$ Nikon (with numerical aperture 0.8) objective. The reflected signal was collected by the same objective and coupled into a fibre. The output from the fibre was analyzed with a Princeton Instruments spectrometer with a nitrogen-cooled charge-coupled device. 

\textbf{Dark-field spectroscopy}: The dark-field spectroscopy was carried out with the same microscope as for the bright-field spectroscopy in the dark-field configuration. White light illumination was focused onto the sample at a high incidence angle of 37$^o$ from the surface normal. Light scattered normally was then collected into a fibre and analysed with the spectroscopy set-up.  

\textbf{Finite-difference time-domain (FDTD) scattering simulation}: The FDTD simulations were carried out by using Lumerical software. The refractive index of WS$_2$ obtained using ellipsometry in Ref. \cite{Zotev2023} was used. The refractive index of gold was from Ref. \cite{palik1998handbook}. The refractive index of the monolayer WSe$_2$ was extracted from Ref. \cite{2019RN406}.

\textbf{Nonlinear RC spectroscopy}: We used a supercontinuum laser (NKT-SuperK Extreme) as the light source which was focussed on the sample with a 100$\times$ objective (0.9 numerical aperture) with a focused beam diameter of 2.3 $\mu$m. The reflected signal from the sample is then collected with the same objective and analyzed with the spectroscopy system.

\section{Acknowledgements}

Yadong Wang (Y.W.) and A.I.T. acknowledges UKRI Horizon Europe Guarantee award for MSCA postdoctoral fellowship TWIST-NANOSPEC. C. L., P.G.Z., S.R., X.H., O.P.C., P. B. and A.I.T. acknowledge support from the European Graphene Flagship Project under grant agreement number 881603 and EPSRC grants EP/S030751/1, EP/V006975/1, EP/V007696/1 and EP/V026496/1. Yue Wang thanks EPSRC grant
EP/V047663/1 and Royal Academy of Engineering fellowship RF/201718/17131.

\section{Author contributions}
Y.W. and A.I.T conceived the research idea. X.H., S.R, O.P., Yue W., C.K.C. and R.G. fabricated the studied samples. Y.W. and S.R. characterised NAs using SEM. Y.W. carried out optical characterization and analyzed the results. Y.W. and C.L. analyzed the nonlinearity of Mie-polaritons. Y.W. simulated the scattering spectra, and electric field distributions with the help of P.G.Z., S.R and P.B. Y.W., C.L., S.R. and A.I.T. wrote the manuscript with contributions from all authors. A.I.T., Yue W. and R.G. provided management for various aspects of the project.

\bibliography{Manuscript_V2}



\hfill \break

\end{document}


\Large

\title{\Large Supplementary information for Exciton-polaritons in a monolayer semiconductor coupled to van der Waals dielectric nanoantennas on a metallic mirror}

\author{Yadong Wang}
\email{yadong.wang@sheffield.ac.uk}
\affiliation{School of Mathematical and Physical Sciences, University of Sheffield, Sheffield S3 7RH, UK}

\author{Charalambos Louca}
\affiliation{School of Mathematical and Physical Sciences, University of Sheffield, Sheffield S3 7RH, UK}

\author{Sam Randerson}
\affiliation{School of Mathematical and Physical Sciences, University of Sheffield, Sheffield S3 7RH, UK}
\author{Xuerong Hu}
\affiliation{School of Mathematical and Physical Sciences, University of Sheffield, Sheffield S3 7RH, UK}

\author{Panaiot G. Zotev}
\affiliation{School of Mathematical and Physical Sciences, University of Sheffield, Sheffield S3 7RH, UK}
\author{Oscar Palma Chaundler}
\affiliation{School of Mathematical and Physical Sciences, University of Sheffield, Sheffield S3 7RH, UK}
\author{Paul Bouteyre}
\affiliation{School of Mathematical and Physical Sciences, University of Sheffield, Sheffield S3 7RH, UK}
\author{Casey K. Cheung}
\affiliation{Department of Physics and Astronomy, The University of Manchester, Oxford Road, Manchester, M13 9PL, UK}
\affiliation{National Graphene Institute, The University of Manchester, Oxford Road, Manchester, M13 9PL, UK}
\author{Roman Gorbachev}
\affiliation{Department of Physics and Astronomy, The University of Manchester, Oxford Road, Manchester, M13 9PL, UK}
\affiliation{National Graphene Institute, The University of Manchester, Oxford Road, Manchester, M13 9PL, UK}
\author{Yue Wang}
\affiliation{School of Physics, Engineering and Technology, University of York, York, YO10 5DD, UK}
\author{Alexander I. Tartakovskii}
\email{a.tartakovskii@sheffield.ac.uk}
\affiliation{School of Mathematical and Physical Sciences, University of Sheffield, Sheffield S3 7RH, UK}


\begin{abstract}

\end{abstract}

\maketitle
\newpage

\setcounter{figure}{0}
\renewcommand{\figurename}{Fig.}
\renewcommand{\thefigure}{S\arabic{figure}}

\begin{figure}[h]%
\includegraphics[width=1\textwidth]{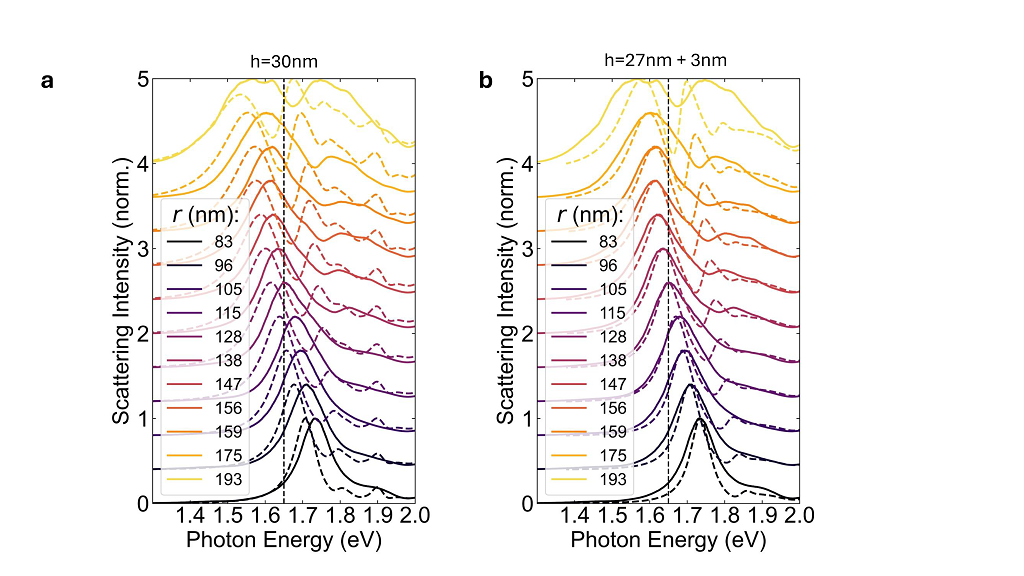}
\caption{\large \textbf{Comparison between simulated and experimentally measured light scattering spectra from WS$_2$ nanoantennas (NAs).}  Experimental (solid lines) and simulated (dashed lines) spectra for the NAs made purely from (\textbf{a}) 30 nm thick WS$_2$ and (\textbf{b}) when the WS$_2$ NAs with thickness 27 nm is placed on a 3 nm tall gold pedestal. The vertical dashed line indicates the energy of the exciton in monolayer WSe$_2$.}
\label{Figure SIM_EXP_Comparison}
\end{figure}
\clearpage

\begin{figure}[h]%
\includegraphics[width=1\textwidth]{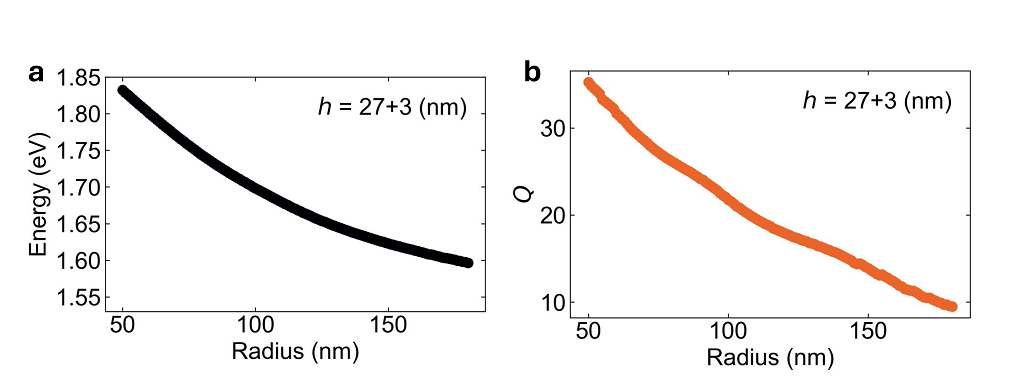}
\caption{\large \textbf{Properties of an electric-dipole mode in WS$_2$ NAs on gold: results of simulations.}  Fitted energy positions (\textbf{a}) and $Q$ factor (\textbf{b}) of the simulated electric-dipole (ED) mode in WS$_2$ NAs on a gold substrate. The NAs have a height of 30 nm combined from a 27 nm WS$_2$ block placed on a 3 nm high gold pedestal.}
\label{Figure WS2 mode fitting}
\end{figure}
\clearpage

\begin{figure}[h]%
\includegraphics[width=1\textwidth]{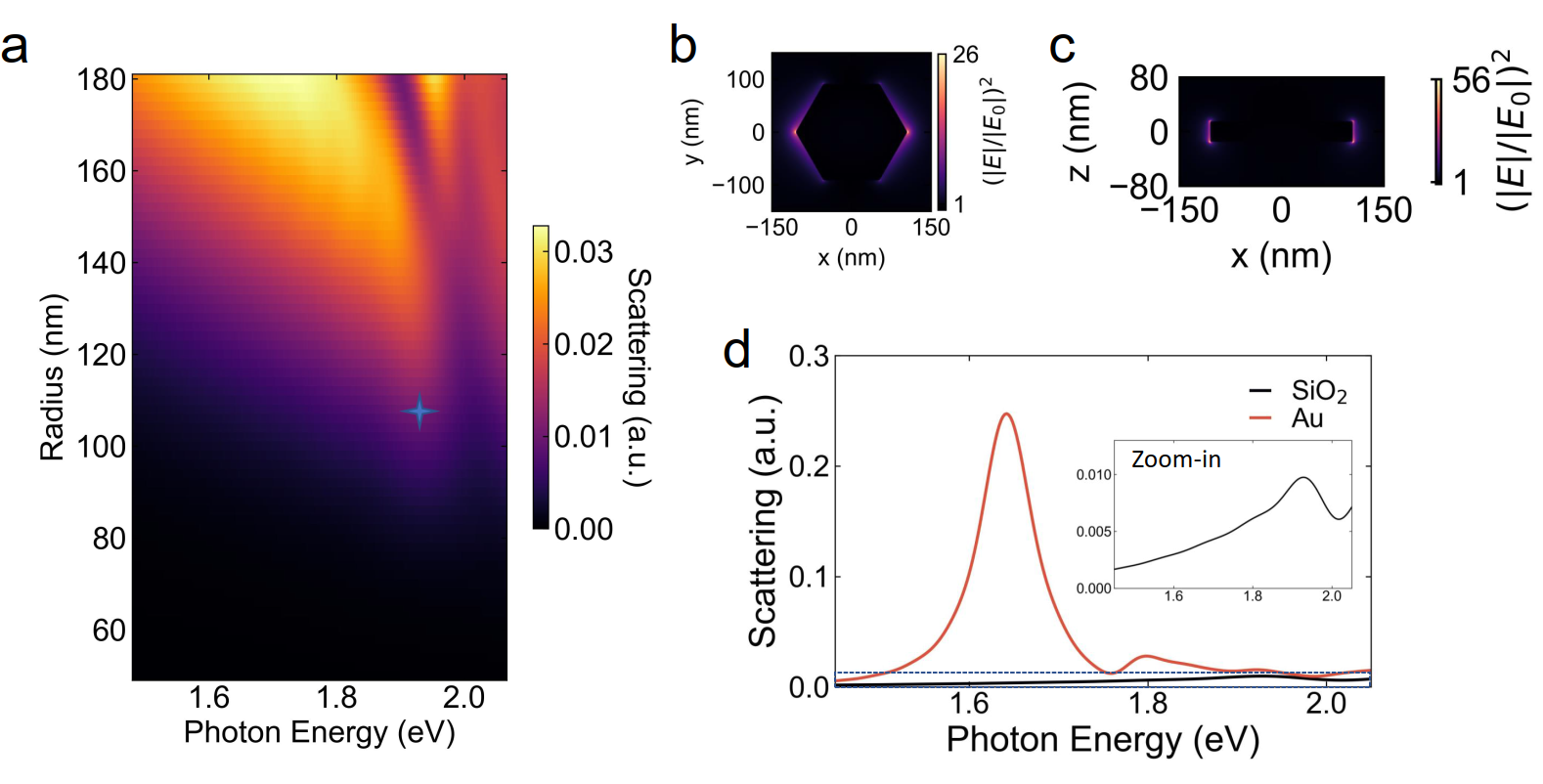}
\caption{\large \textbf{Simulated properties of WS$_2$ NAs on SiO$_2$. } \textbf{a,} Normalized scattering spectra of the WS$_2$ NAs with the thickness of 30 nm and radius ranging from 50 nm to 180 nm with a step of 2 nm. The $xy$-plain (\textbf{b}) and $xz$-plain (\textbf{c}) electric-field distributions for a NA with a radius of 105 nm (star in Fig a) at 638 nm (1.94 eV). \textbf{d, } Comparison of the scattering spectra between the NA on gold and on SiO$_2$. Inset: zoomed-in spectrum of the NA on SiO$_2$.}
\label{Figure WS2 on SiO2}
\end{figure}

\clearpage

\begin{figure}[t]%
\includegraphics[width=1\textwidth]{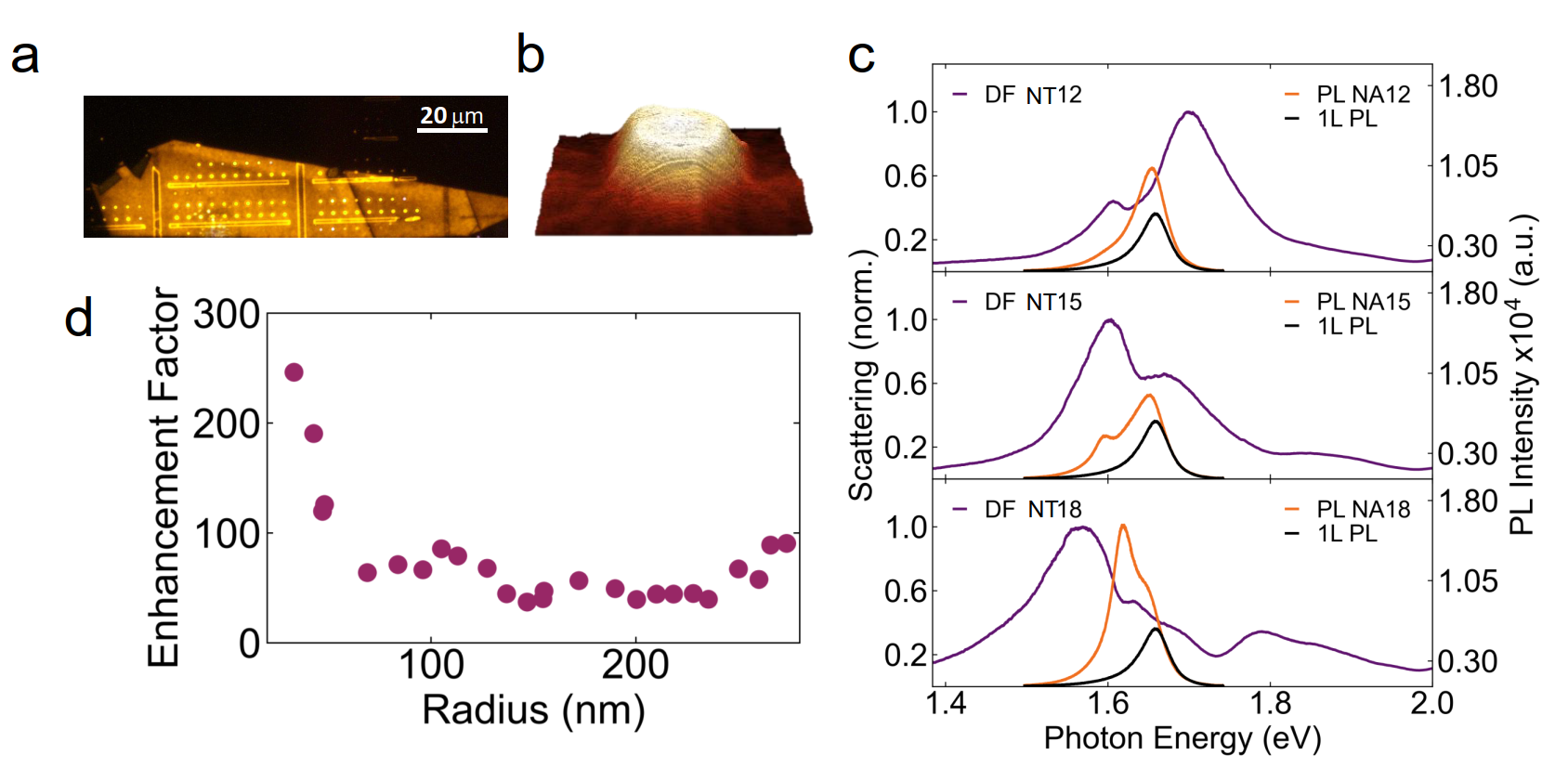}
\caption{\large \textbf{PL enhancement in 1L WSe$_2$ placed on WS$_2$ on gold} \textbf{a,} PL image of an NA array covered by a monolayer WSe$_2$. \textbf{b,} AFM image of an NA covered by the monolayer WSe$_2$. \textbf{c,} Examples of PL spectra of monolayer WSe$_2$ (1L WSe$_2$) on WS$_2$ NAs (orange) plotted together with a PL spectrum measured for 1L WSe$_2$ on gold (black) and a normalized dark-field scattering spectrum. \textbf{d,} PL enhancement on the NAs compared with the monolayer PL on gold calculated taking into account the reduced area of the NA's top surface compared with a laser spot on a planar surface.}
\label{Figure Purcell enhancement}
\end{figure}

\section{Photoluminescence (PL) enhancement in the WSe$_2$ monolayer coupled to WS$_2$ nanoantennas on gold}\label{sec_S2}
Figure \ref{Figure Purcell enhancement}a shows the PL image of the 1L WSe$_2$ covered NA regions. The brighter emission at the NA sites is the indication of the enhanced photoluminescence (PL) from 1L WSe$_2$ due to the coupling with the NAs. The AFM image of one NA confirms that 1L WSe$_2$ tightly wraps onto the surface of NA, as shown in Fig. \ref{Figure Purcell enhancement}b. To evaluate the photonic response of the NAs, we subsequently carried out detailed room-temperature PL measurements in a micro-PL setup. Excitation with a continuous-wave laser at the wavelength of 682 nm below WS$_2$ absorption edge was chosen. Figure \ref{Figure Purcell enhancement}c displays the PL spectra measured from a planar portion of the WSe$_2$ monolayer on gold (black lines) and specific NAs with radii $r_{NA12}$ = 96 nm, $r_{NA15}$ = 127 nm, $r_{NA18}$ = 155 nm (orange). The dark-field spectra from the corresponding NAs are also displayed in the figure for comparison. The red-shift of the PL spectra on the NAs observed here is partly due to the strain in the monolayer as it conforms to the NA shape. We defined an experimental enhancement factor $\langle$EF$\rangle$ by considering the effective excitation area of NAs: 

\begin{equation} \label{eq1}
\begin{split}
\langle EF \rangle & = \frac{I_{NA}-I_{Au}}{I_{Au}} \frac{A_{laser}}{A_{NA}}\\
\end{split}
\end{equation}

Where $I_{NA}$ and $I_{Au}$ are the spectrally integrated PL intensities measured on each NA and on the flat Au substrate, respectively. The area of each NA and the area of the laser spot are represented as $A_{NA}$ and $A_{laser}$, respectively. Using this definition, we calculated the enhancement factors of 246, 126, and 66 for NA12, NA15, and NA18, respectively. Figure \ref{Figure Purcell enhancement}d displays the enhancement factor as a function of the NA radius.

\clearpage

\begin{figure}[t]%
\includegraphics[width=1\textwidth]{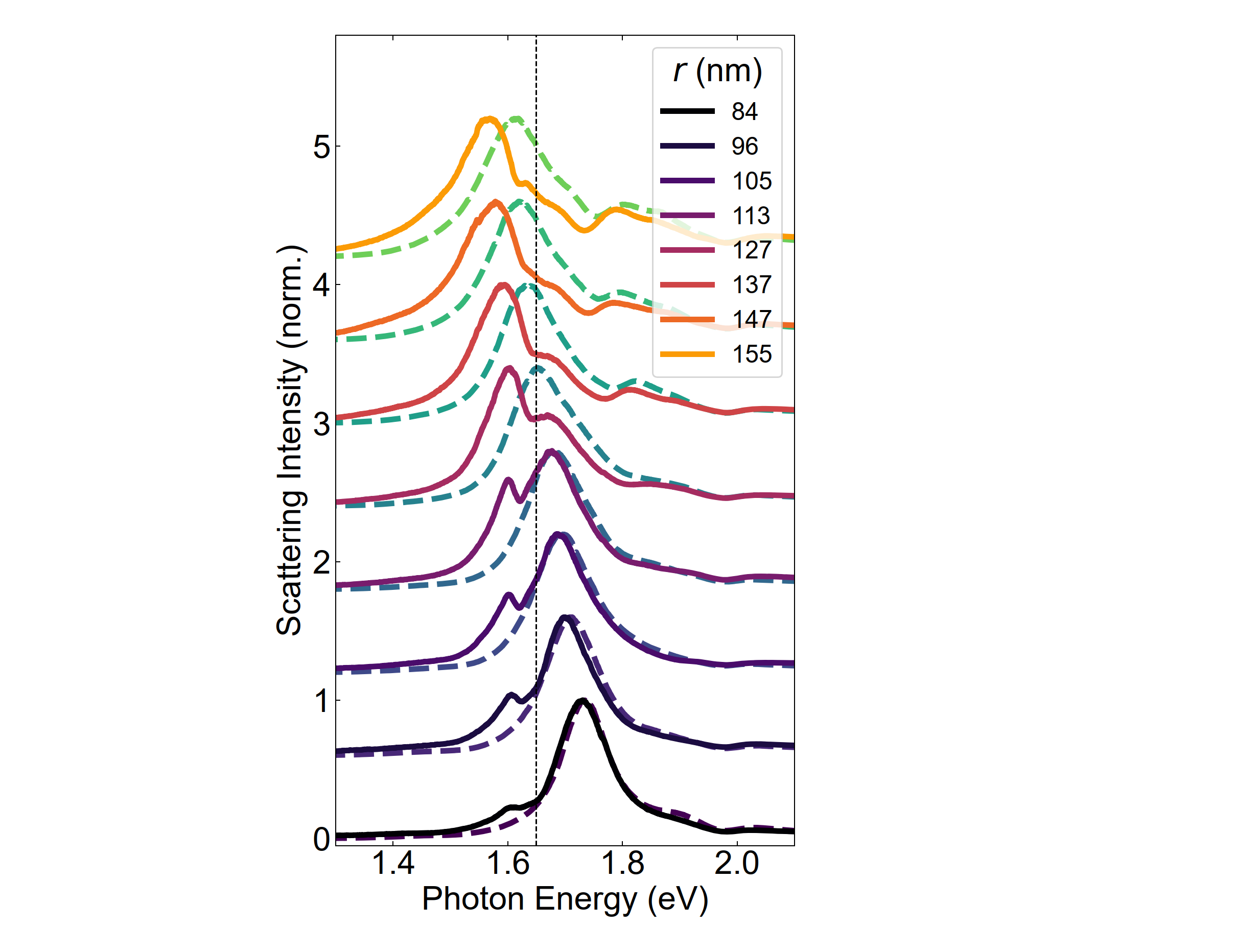}
\caption{\large \textbf{DF spectra comparison before and after transfer.} The green sequential dashed curves show the DF spectra of bare NAs, while the inferno sequential curves show the DF spectra of 1L WSe$_2$ wrapped NAs. The black dashed line indicate position of exciton energy and two dashed grey curves donate the upper and lower polaritons, respectively.}
\label{transferbeforeafter}
\end{figure}

\clearpage

\section{Coupled Oscillator Model}\label{sec3}

We further evaluated the strength of light-matter interactions with exciton and Mie resonance with the dark-field scattering (DF) spectra as a function of different NA radii. We observed that a characteristic anti-crossing arises as the Mie mode is tuned in resonance with WSe$_2$ exciton. To map the polariton dispersion curve as a function of the NA size we fit the two peaks in the DF spectra with Lorentzians. The results of the fitting are presented in Fig.3 of the main text. We observe an anticrossing behaviour typical for the strong coupling between the exciton in the WSe$_2$ monolayer and the Mie-resonance. 

Exciton-polaritons arise as a result of the strong coupling between the exciton and photonic mode observed when their respective dissipation rates are slower than the coupling rate between the two states \cite{Kavokin2017}. These part-light part-matter quasi-particles can be realised via exciton coupling with a variety of photonic modes in optical microcavities, photonic crystals, metasurfaces, nanoantennas, waveguides etc. 

The effect of the exciton-photon coupling is dependent on the energy detuning between the photonic mode and the exciton. For a given exciton energy $E_X$, the detuning can be changed by varying the geometrical parameters of the NA resulting in the change of the photonic mode energy $E_{ph}$.  In our case, the photonic mode dispersion $E_{ph}$ can be expressed as a function of the NA radius $r$, while the exciton energy $E_X$ can be considered constant.

The coupling between the exciton and the photonic mode can be described with a two-level model. In this model, the dissipation rates of the two states are given by their respective linewidths, $\gamma_{X}$ and $\gamma_{ph}$. The coupled states are retrieved by solving the following Hamiltonian:

\begin{equation}
\label{eq_Hamiltonian}
H= 
\begin{pmatrix}
E_{ph}(r) -i\gamma_{ph} & g \\
g  & E_{X} -i\gamma_{X}
\end{pmatrix},
\end{equation} 

\noindent in which $g$ is the coupling strength between the exciton and the photonic mode depending on the exciton oscillator strength $f_X$, the permittivity of the material $\epsilon_r$, and the photonic mode volume $V$:

\begin{equation}
\label{eq_Rabi_energy_2}
\centering
g=\sqrt{\frac{e^2}{4\pi\epsilon_0\epsilon_r}\frac{f_{X}}{V}}\propto\sqrt{\frac{f_{X}}{V}}.
\end{equation}

The eigenstate energies $E_\pm$ and linewidths $\gamma_\pm$ are given by the real and imaginary parts of the complex eigenvalues of the system of the coupled states: 

\begin{equation}
\label{eq_eigenvalues}
\centering
E_{\pm}(r)=\frac{1}{2}[E_{ph}(r)+E_{X}-i(\gamma_{ph}+\gamma_{X})]\pm\frac{1}{2}\sqrt{[E_{X}-E_{ph}(r)+i(\gamma_{ph}-\gamma_{X})]^2+4g^2}.
\end{equation} 

The Rabi splitting of this system, $\hbar\Omega$, defined as the energy splitting between the two eigenstates at the resonance of the exciton and photonic states when $E_{ph}(r)=E_X$, is given by: 

\begin{equation}
\label{eq_Rabi_energy_1}
\centering
\hbar\Omega = \sqrt{4g^2-(\gamma_{ph}-\gamma_X)^2}.
\end{equation}

When the Rabi splitting is a real number, i.e when $g\ge\frac{|\gamma_{ph}-\gamma_X|}{2}$, the eigenenergies $E_\pm$ split and the system is in the strong coupling regime. In this case, the two eigenstates are exciton-polaritons characterized by an anti-crossing at the crossing point of the uncoupled exciton and photonic dispersions as observed in our case in Fig.3 of the main text. 

As the observation of the characteristic polariton anti-crossing is required experimentally, the Rabi splitting should also be larger than a half of the sum of the photon and exciton linewidth, i.e. $\hbar\Omega\ge(\gamma_{ph}+\gamma_X)/2$, fulfilled in our case.

\clearpage

\begin{figure}[t]%
\centering
\includegraphics[width=1\textwidth]{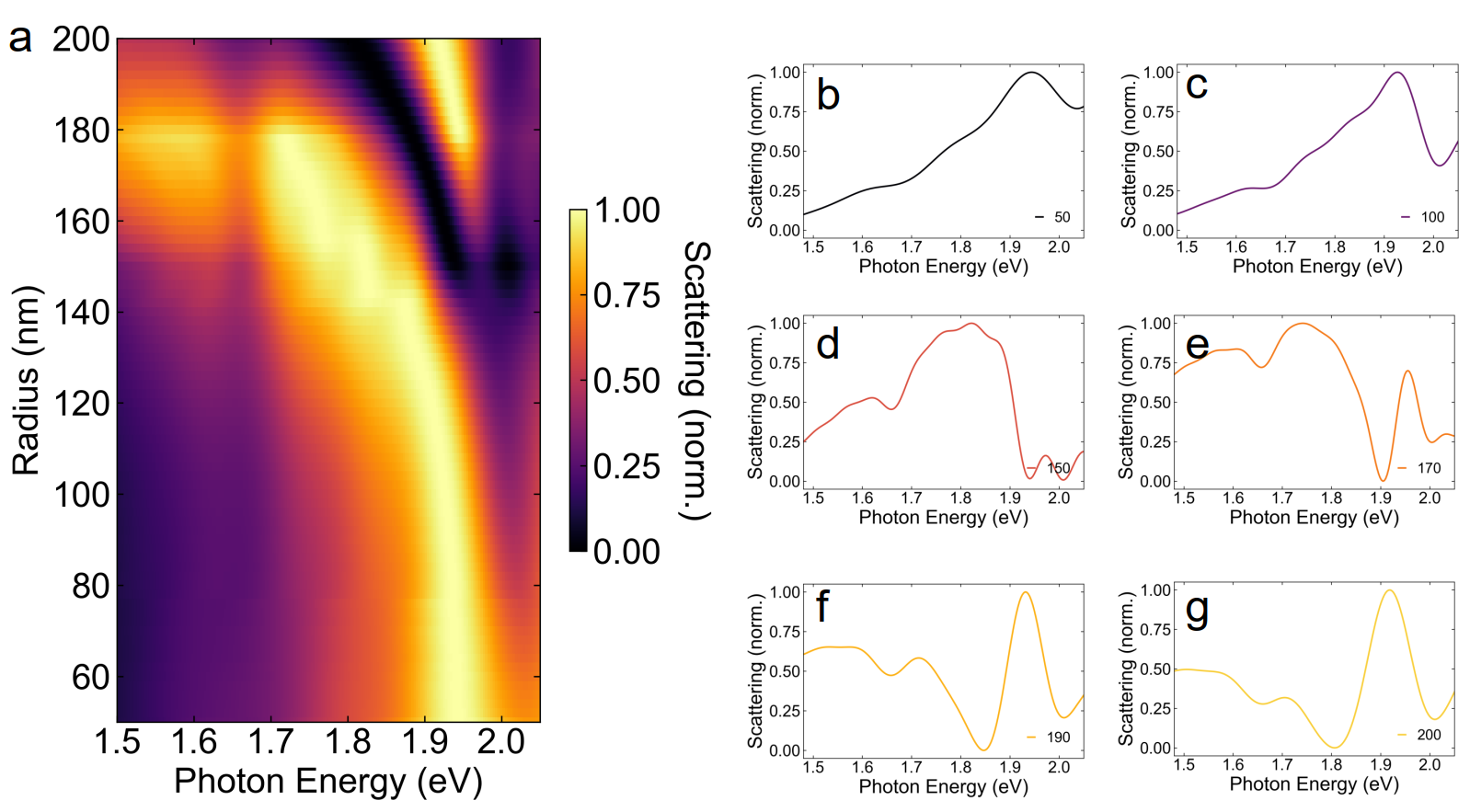}
\caption{\large \textbf{Simulated scattering spectra in the WS$_2$ NAs enveloped with a WSe$_2$ monolayer and placed on a SiO$_2$ substrate.} \textbf{a,} Normalized scattering spectra of the WS$_2$ NAs with the thickness of 30 nm and radii ranging from 50 nm to 200 nm with a step of 2 nm on a SiO$_2$ substrate. \textbf{b-g} Individual scattering spectra for radii of 50, 100, 150, 170, 190, 200 nm. It is notable that the strong coupling observed for Mie resonances in the WS$_2$ NAs on gold covered with a WSe$_2$ monolayer is not present in the case of a SiO$_2$ substrate.}
\label{Figure WSe2-NA on SiO2}
\end{figure}

\begin{figure}[t]%
\includegraphics[width=1\textwidth]{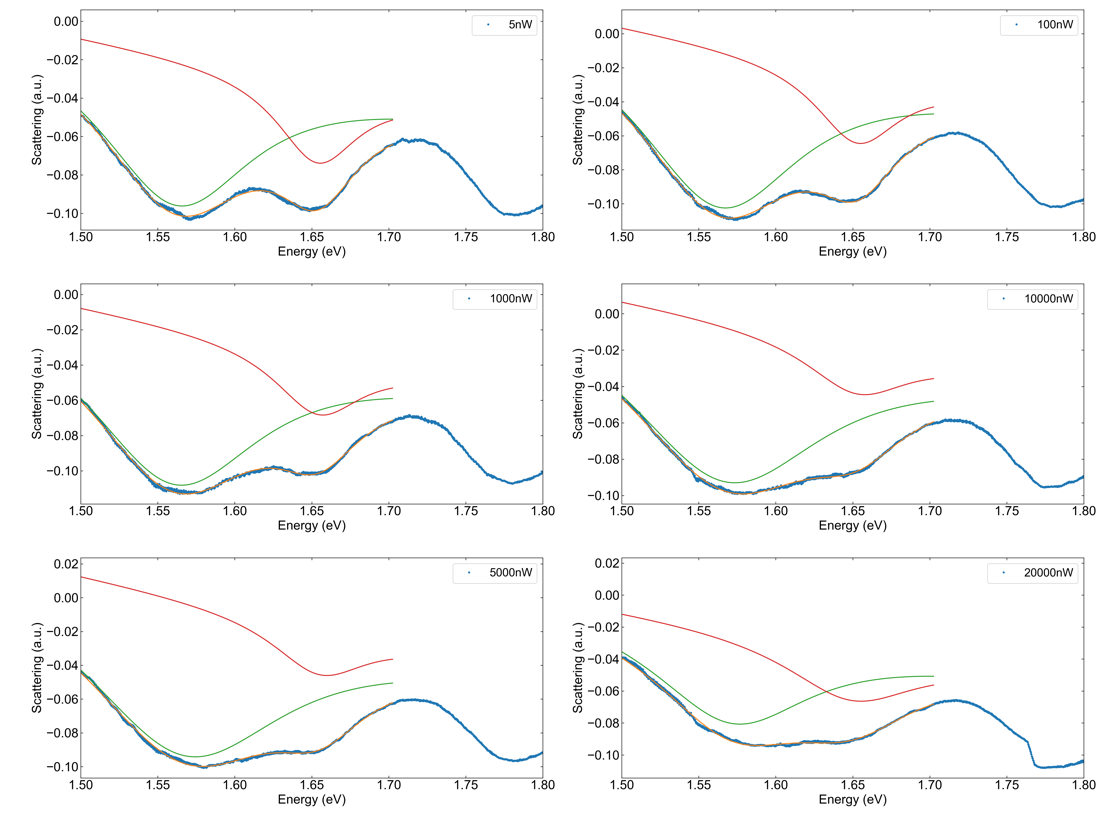}
\caption{\large \textbf{Fitting of RC spectra for selected excitation powers of the resonant excitation with a supercontinuum laser.} Bi-Lorentzian fitting of the RC spectra at different powers from 5 nW to 20 $\mu$W.}
\label{Figure fitting powerdep}
\end{figure}
\clearpage

\bibliography{Manuscript_V2}

\hfill \break